\documentstyle [12pt,a4wide,epsf]{article}
\begin{document}

\thispagestyle{empty}
\setcounter{page}{0}

\flushright{Imperial/TP/93-94/23}
\vspace{0.25in}

\begin{center}
\begin{bf}
\begin{Huge}
Cosmic Rays From Cosmic Strings\\
\end{Huge}
\end{bf}
\vspace{0.5in}
A.J. Gill and T.W.B. Kibble\\
\vspace{0.25in}
Blackett Laboratory\\
Imperial College\\
South Kensington\\
London SW7 2BZ\\
United Kingdom\\
\vspace{0.25in}
February 1994\\
\end{center}
\vspace{0.5in}

\abstract{It has been speculated that cosmic string networks could
produce ultra-high energy cosmic rays as a by-product of their
evolution. By making use of recent work on
the evolution of such networks, it will be shown that the
flux of cosmic rays from cosmologically useful, that is GUT scale
strings, is too small to be used as a test for strings with any
foreseeable technology.}

\newpage

\section{Introduction}

Cosmic strings produced during GUT scale phase transitions in the
early universe could provide a mechanism for cosmological structure
formation \cite{struct}. They also preserve regions of space-time in
the symmetry unbroken phase due to boundary conditions which
topologically restrict their decay. Under certain circumstances,
these restrictions are removed and the energy stored in the unbroken
vacuum phase is liberated in the form of the GUT scale quanta of the
gauge and scalar fields which form the defects. It has been suggested
that the subsequent decay of these massive GUT quanta could be a
source of ultra-high energy cosmic rays \cite{bhatt1} both at and
above the highest energies currently observable. If enough of these
cosmic rays are produced then they could  provide a clean
observational signal characteristic of cosmic strings.

Such a scheme for producing ultra-high energy cosmic rays obviates the
need for acceleration mechanisms, such as those involving AGNs, which
are currently thought to be responsible for most of the highest energy cosmic
rays observed \cite{agn}. Thus, high energy {\it neutral} as well as
charged particles could be produced, although the radiation from
cosmic strings would not contain any nuclei heavier than hydrogen.
Also, acceleration mechanisms in
AGNs are thought to be incapable of producing cosmic rays which would
reach us at energies above $10^{19}$eV.  Thus, if the flux of
ultra-high-energy
cosmic rays were large enough to be measurable, we would have a very
clean, background free test for cosmic strings. Even knowing that
strings are not capable of producing
cosmic rays at such high energies is advantageous as anything which is
then seen must be from some previously unknown process and therefore
indicate the presence of entirely new physics. Experiments are now
either running, being built or being designed to measure the ultra-high energy
cosmic ray flux at appropriate energies \cite{flyhaverah} \cite{dumand}.

Prior to this, much work had already been done on the mechanisms by
which strings decay to the particles commonly found in cosmic rays
\cite{sredthei} and also on the
propagation of high-energy cosmic rays \cite{prop}. Due to calculational and
computational difficulties concerning the evolution of cosmic string
networks, however, very little predictive work had been done on the likely flux
of cosmic rays and no work at all had been
done which included all possible mechanisms for removing topological
stability. Thus the magnitude of the expected cosmic ray flux was a
completely unknown quantity. In much previous work, the predictions
were simply normalised to match the experimental cosmic ray data at around a
joule. Despite the lack of a solid justification, no criticism was made of such
techniques and the predictions
were adopted uncritically by the astrophysics community \cite{calgary}.
In addition to this, our understanding of some of the
assumptions made in the early work has changed significantly since
this was done, to the extent that some of the earlier results must now
be modified.

It is also possible that other topological defects could
produce observable cosmic rays, but this is as yet a virgin field as
our ideas of the evolution of other defects are not as advanced as
those concerning strings.

\section{The Mechanism Of Cosmic Ray Production}

Cosmic strings, and topological defects in general, are soliton solutions
of classical equations which, it is claimed, are the effective field
equations of the, as yet unknown, GUT-scale physics \cite{gvh}
\cite{langacker}. These
solutions describe the vacuum expectation value of a scalar field
$\langle \phi \rangle$ and its associated gauge field. The scalar field
involved may thus be either a genuine scalar, such as the Higgs, or a
composite structure somewhat like a Cooper pair. Such defects
may be formed in gauge theories during phase transitions in which the
full symmetry group is broken in some fashion
\cite{twbkorig} \cite{twbkrev}.

One of the advantages of
the cosmic string model is the paucity of free parameters. All of the
{\it cosmological} consequences of strings are relatively insensitive
to the coupling
constants of the field theory and depend almost entirely on a single
parameter related to the temperature of the symmetry-breaking phase transition.
This parameter is usually specified in the form of the dimensionless
number $G \mu / c^4$ where $\mu$ is the energy per unit length of the
string. For cosmologically useful strings, this would have to be of
the order of $10^{-6}$. The cosmic ray flux also depends in
principle not just on $\mu$ but also on the values of the coupling
constants, although in practice
this is not a problem as one can place a unitarity bound on them and
thence obtain an upper limit on the flux.

Once formed, strings preserve linear regions of space-time in the
symmmetry unbroken phase since the boundary condition that the fields
should be in their symmetry broken state at infinity places a
topological restriction on their evolution. Thus, a network of cosmic
strings stores a large amount of energy. How the network evolves
determines the fate of this energy.

The string network may be described statistically by various physical
length scales, for example the characteristic distance between
strings. Such scales are time dependent, their time evolution
describing how the network changes with time. The network can evolve in one
of two possible ways. Causality prevents the physical length scales growing
faster than $ct$ so either the length scales are all proportional to
$ct$, which is referred to as scaling, or the length scales grow less
fast than $ct$ and the strings come to dominate the energy density of
the universe \cite{scaling}. Which of these two alternatives describes
a real string
network depends on whether the network can lose energy fast enough to
scale.\footnote{It is worth pointing out that the string networks observed in
nematic liquid crystals do indeed seem to scale \cite{nematic}.} If
strings are to be a viable cosmological model then they must scale as
string domination is ruled out by observation. The energy lost from
a cosmic string network provides us with a way to look for strings and
thus probe some of the GUT scale physics which would be responsible
for them.

There are two ways in which cosmic strings can lose energy.
Firstly they can radiate gravitationally, which provides one of the
current bounds on $G \mu / c^4$. Alternatively, one can remove the
topological stability from a length of string in some way. This is the
basis of the mechanism whereby strings can produce cosmic rays.
There are three possible ways of doing this, viz. inter-commutation,
loop collapse and cusp evaporation.
To calculate the expected flux of cosmic rays, it is necessary to know
how many of each of these types of event occur at each epoch, each
event producing quanta of either the scalar or gauge fields.
Previously, the most general work had assumed a generic
two parameter form for the event rate and thus the number of GUT
quanta produced at each epoch. This was based on the idea of
scaling \cite{bhatt2}. However, the
two parameters involved do not have obvious physical meanings.  In the
light of recent analytical work \cite{ted}, it is possible to improve
on this generic fit and write expressions involving only physically
meaningful parameters whose values and uncertainties are relatively
well known from simulations.

GUT and gauge quanta are therefore produced and decay rapidly on
moving through the symmetry broken phase outside the string. Whatever
the actual GUT
responsible for the physics, the final decay products must be stable
particles with which we are familiar. Whatever the high energy
physics, there is ultimately a very restricted range of decay routes once
at the relatively low energies with which we are familiar. Three
generic decay routes stand out. Firstly, the decay may end up
producing gauge bosons; these do not
propagate far as they either decay, like the Ws or Z, or interact with
the inter-galactic medium. They tend to interact and produce other
particles. A leptonic
route tends to produce a large number of neutrinos and electrons, with
a few hadrons and heavier charged leptons. These neutrinos could be
important because of their comparatively long path lengths in the
inter-galactic medium. Finally there is the QCD decay route in which a
jet is formed from a quark-gluon source.

The theory of this scheme has already been extremely well studied
\cite{bhatt3}. It is the QCD jets which arouse the most interest as they
produce a relatively large number of particles. They also have the
advantage that there is a clear dichotomy between the observations
which are characteristic of decay from GUT scale energies and those
which are characteristic of the string network itself. The energy
spectrum of the particles from a single jet depends almost exclusively
on the QCD and is therefore characteristic of decay from a GUT
particle, whereas the statistics of the jets
themselves and the event rates, if they could be observed, would be
characteristic of the particular defect responsible. On the other
hand, if single
high-energy particles were to be observed then it would be almost impossible
to say conclusively whether these were from cosmic strings or from
some other as yet unknown process involving decay from GUT energies.

Once one has an injection spectrum of ultra-high-energy particles, the
propagation of these particles is a relatively well understood
problem \cite{cosmo} \cite{prop}, although it turns out that it is
not really worth doing this part of
the calculation since the event rate is so low.

\section{Calculation Of The Cosmic Ray Flux}

As stated previously, there are three types of events which remove
topological stability from a length of string, viz. intercommutation,
loop-collapse and cusp formation. All of these present difficulties
for either numerical or analytic treatments of the field theory.
In principle, a sufficiently
good computer simulation would solve all of these problems, up to the
freedom of choice of the two coupling constants and the symmetry
breaking scale. In practice, however, such an approach is not tractable
with current technology, mostly due to the difficulty of obtaining a
sufficiently great dynamic range.

Consider an intercommutation event in which two lengths of string
collide and swap partners \cite{gvh}.
Simulations, particularly of global strings,  lead one to expect that
energy is lost during the exchange. Physically, one expects this to be
of the order of $\mu w$, where $w$ is the width of the strings and to be
in the form of GUT scale gauge and scalar bosons. Henceforth these
will be referred to collectively as X-particles; this terminology does
not imply that they are GUT gauge bosons and is adopted merely
to be consistent with other papers in the field. The number of
X-particles produced per intercommutation event will be $(\mu
w) / (m_{X}c^{2})$, where $w \approx 10^{-31}$m, is the width of the
string, and $m_X$ is the mass of the GUT quanta. This is roughly
of order one, within an order of magnitude or so, depending on exactly
how one fixes $\mu$ and $m_X$.

Loops of cosmic string are formed when a length of string intersects
with itself; intercommutation then chops off a loop of string which
begins evolution independently of the original length. Loops of cosmic
string normally radiate almost entirely gravitationally, shrinking as
they do so. As a loop shrinks, it will probably chop itself up into
smaller loops by inter-commutation events.
Ultimately, loops reach a point when their diameter is of the order of
the string radius and their topological stability is removed by field
overlap. At this point, the energy remaining in the string is
liberated as X-particles and the number of particles is again of the
order $\mu w / m_{X}c^{2}$, neglecting factors of $\pi$.

It is not known how many inter-commutation events are involved in chopping
up a loop prior to its demise or indeed when the loop `realises' that it
is unstable. This uncertainty may introduce a factor of a thousand or so but
the important point is that the number of X-particles produced by the
demise of a loop is fixed. It is not time dependent. Previously, it
was sometimes assumed that a certain fixed fraction of the energy initially
in a loop would end up as cosmic rays. Such an assumption does not
accurately reflect the physics, however, as it would imply that the
energy liberated in each loop demise scales since the average loop size
is directly proportional to time. Instead of growing with time, the
energy liberated per event is actually constant.

Cusps are points where the radius of curvature of the string becomes
very small. Mathematically, the radius of curvature is undefined
for a cusp but backreaction processes prevent this ever quite
happening. The string may be approximately described by the Nambu-Goto
action, from which it is possible to deduce that {\it instantaneously} a
cusp moves along relative to the string at the speed of light.

Cusps have always been thought to radiate massive gauge and scalar
bosons, which is a part of the backreaction mentioned above. According
to a recent calculation by Mohazzab \cite{mozzie}, the energy radiated
by a cusp is of the order of
$\frac{1}{2}g^{5/2} \eta$ where $\eta$ is the
symmetry breaking scale and $g$ is the coupling constant between the
string scalar field and whatever is radiated. This is much smaller
than was previously held to be the case. Broadly speaking this is
because the sizes of the incoming waves are
much bigger than the thickness of the cosmic string.  Choosing $g$ to
be close to the unitarity limit, that is to be about one, would give a
plausible upper bound on the cosmic-ray flux from cusps. However, this
perturbative result is odd in that it is totally independent of the
shape of the cusp and it might be suspected that a higher order
calculation would bring to light some small dependence of this sort,
which would increase the energy released.
To be conservative in the upper limit on the flux, therefore, old
ideas about cusps will be employed which give much larger,
time-dependent fluxes.
According to the old ideas of cusps, the length of string which
overlaps in a cusp is of the order of $(\zeta^2 w)^{1/3}$, where
$\zeta$ is the characteristic length of the small-scale structure on
strings; essentially the interkink distance. Such a length scales, so
one may write $\zeta = c/ \epsilon H = 3ct / 2 \epsilon$ in the matter
era where the parameter $\epsilon$ is dimensionless. This
suggests that the number of X-particles radiated will be something
like $(3ct/2 \epsilon w)^{2/3}$. For the sake of a reliable upper
bound, this is the expression we will use, rather than that of
Mohazzab, although the latter may well be more accurate.

If we wish to calculate a cosmic ray flux then in addition to knowing
the energy produced per topological stability removing event, one also
needs some idea of how many such events there are likely to be and
therefore some idea of how the network evolves. There exist three
major simulations \cite{as} \cite{at} \cite{bb} and one analytic study
\cite{ted}, although these only give a
rough indication of the evolution since this is a very difficult problem. A
combination of these enables one to say something about the number of
stability removing events and therefore the rate
of production of X-particles; the analytic being used to extract the
relevant physics from the simulations without the restrictions on the
dynamic range. In what follows, much of the notation is the same as
that used in the analytic study \cite{ted}.

Consider first intercommutations. The length of string in a volume $V$
is $L=V/ \xi^2$, where $\xi$ is a
length scale related to the overall density of strings. The
probability that such a length intercommutes per unit time is $p=
\chi L c / \xi^2$, where from the simulations $\chi$ is probably about
$0.03$ and certainly no larger than $0.2$.
Thus the expected number of intercommutations per unit volume is:-
\begin{displaymath}
\langle n_{ic}(t) \rangle = \frac {\chi c}{\xi^{4} }.
\end{displaymath}
Following the same procedure as the aforementioned study, we
define a dimensionless constant $\gamma = c/ H \xi$ where in the
matter-dominated era $\gamma$ is somewhere
between one and four and $H=2 / 3t$, thus giving:-
\begin{displaymath}
\langle n_{ic}(t) \rangle = \frac{\chi (H \gamma)^{4}}{c^3} = \frac
{16 \chi}{81 c^3}.
\biggl ( \frac{\gamma}{t} \biggr ) ^{4}
\end{displaymath}

Loops are a better understood problem since they were once thought to
be the crux of a successful structure formation model. Once formed,
loops of cosmic string radiate energy in the form of gravitational
waves until they reach the point when the overlap between the field
on different sides of the loop causes the loop to realise that it is
not topologically stable. This gives loops a life-time of
$l_b c^3 / \Gamma G \mu$, where $l_b$ is their initial length and
$\Gamma$ is a dimensionless constant of order $100$, governing the
rate of gravitational radiation.

Let $l_b(t_b)$ be the mean length of loops born at $t_b$.
For simplicity, we shall assume that all loops are born with length
$l_b$. Their length at a later time $t$ will then be $l = l_b - \Gamma
 G \mu ( t-t_b )$. It is convenient to define a dimensionless constant
$K$ by setting:-
\begin{displaymath}
l_b=(K-1) \Gamma G \mu t_b / c^3.
\end{displaymath}
It then follows that the loops born at $t_b$ shrink by the emission of
gravitational radiation and finally expire at time $t=Kt_b$.

How large is $K$? In the analytic study mentioned above \cite{ted}, it
was concluded that the characteristic length of the small-scale
structure on strings, $\zeta$ is likely to be smaller than $\xi$ by a
factor of order $\Gamma G \mu / c^4 \approx 10^{-4}$. Setting
$\epsilon = c/ H \zeta$, as above, $\epsilon \approx 10^4$. Moreover,
the simulations suggest that most loops are born with a length much
smaller than $\xi$, of the order of a few times $\zeta$. It follows
that we should expect $K$ to be of the order of a few, probably betwen
$2$ and $10$ say.

We also need to know how many loops are formed. The number of loops
decaying to X-particles per unit time per unit volume at a time $t$ is
equal to the number born at a time $t_b = t/K$ in a volume $\bigl (
 a(t_b) / a(t) \bigr ) ^{3}$. Now
the number of loops born at a time $t$ in a volume $V$ is given by:-
\begin{displaymath}
\dot{N} = \frac{\nu V c}{(K-1) \Gamma G \mu t_b^4}.
\end{displaymath}
The parameter $\nu$ is dimensionless and characterises the rate of
loss of energy in long strings by loop formation \footnote{It is related
to the parameters defined in \cite{ted} by $\nu = 8f c_0
\overline{\gamma} \gamma ^2 / 27$, where $f \approx 0.7$ is the
fraction of the loop energy that goes into gravitational radiation
\cite{alandcold} and $c_0$
denotes the parameter designated $c$ in the analytic study. Using these
values, we have $c \overline{\gamma} \approx 0.1$,
whence $\nu$, also of order $0.1$.}.
Thus the number of loops expiring per unit volume at a time $t$ which
were born at a prior time $t_b$ is:-
\begin{displaymath}
\langle n_l (t) \rangle  = \frac{\nu c}{(K-1) \Gamma
G \mu} \Biggl [ \frac{a(t_b)}{a(t)} \Biggr ] ^3  t_b^{-4}.
\end{displaymath}
In this approximation where all loops are born with a length equal to
the average length of a loop at birth, all of the loops born at the
same time die at the same time and
vice-versa so to find the total number of loops expiring at a
particular time, there is no need to integrate over birthdays. Also, in
the matter dominated era, $a \propto t^{2/3}$, so one obtains the
final result:-
\begin{displaymath}
\langle n_l(t) \rangle = \frac{K \nu c}{(K-1) \Gamma G \mu t^4}.
\end{displaymath}

Cusps are the least well understood of the possible sources of cosmic
rays, although there has been at least one convincing study
\cite{bran1} \cite{bran2}. In
a scaling regime, the number of cusps per unit time and length on a
long string must be
given by one over the square of some length scale, for example $c/
\overline{\xi}^2$, where $\overline{\xi}$ is the correlation length
along the string. However, it is far from certain that
$\overline{\xi}$ is the right scale to use. Arguably, cusps have more
to do with the small-scale structure, governed by the smallest of the
length scales, $\zeta$, which is smaller than $\overline{\xi}$ by a
factor of order $\Gamma G \mu / c^4 \approx 10^{-4}$. In order to obtain an
upper limit on the event rate the smallest length scale, $\zeta$ will
be used.

If $C^{LS}$ and $C^{L}$ are the numbers of cusps created per
unit length per unit time on long string and loops respectively, then:-
\begin{displaymath}
\langle n_c (t) \rangle = L_{LS} C^{LS} + L_{L} C^{L},
\end{displaymath}
where $L_{LS}$ and $L_{L}$ are the lengths of string per unit volume
in long string and in loops respectively.
If one assumes that, on long string, the number of cusps per unit
length per unit time is $C_{LS} = f_{LS} c/ \zeta^2$, then $f_{LS}$
will be a dimensionless constant at most of order unity. As before,
the length of
long string per unit volume is $L_{LS} = 1 / {\xi}^2 $, where $\xi = c
/ H \gamma = 3ct / 2 \gamma$ in the matter dominated era.  Similarly
one may assume $L_{L} C^{L}$ to be $f_{L} c/ \zeta^2 \xi^2$. Defining
the constant $f=
f_{LS}+f_{L}$, and exploiting the fact that during the matter era $H =
2/3t$, the number of cusp events per unit volume per unit
time becomes:-
\begin{displaymath}
\langle n_c (t) \rangle = \frac {16 f \epsilon^{2} \gamma^{2} }{81
c^{3} t^{4} }.
\end{displaymath}

It will be seen that this also goes like $1/t^4$ as previously.
There is in fact a general argument as to why such event rates should go
like $1/t^4$. One expects $dn$ to be proportional to $V$ and $dt$. The
only dimensionful quantities one can put in the denominator to make
the dimensions correct are lengths or times. Since all of the relevant and
physically meaningful lengths scale \footnote{A priori one might
expect the width of the string, which does not scale,  to be relevant
to cusps but with the optimistic assumptions made this is not the case.},
the denominator must go like $t^4$.

It will now be shown that a calculation for the propagation of the
rays is
superfluous as the flux is far too small for spectra to be useful.
Assume that the beam of cosmic ray particles from each
X-particle event does not disperse too widely and counts as
one event if observed. Also assume that all of the detected events
are useful, which may not be the case with single particles. It turns
out to be clearest to work with conformal time and space co-ordinates
$\tau$ and $x$ as used in the metric for a spatially flat
Friedmann-Robertson-Walker universe:-
\begin{displaymath}
ds^2 = a^2(\tau)(c^2 d\tau^2 - dx^2).
\end{displaymath}
In what follows, a zero subscript will always refer to the value of a
quantity at the present day.

The number of events produced in a spherical shell between $x$ and
$x+dx$ during a conformal time interval $\tau$ to $\tau+ d\tau$ is of
the form:-
\begin{displaymath}
\langle n_t(t) \rangle \, 4 \pi a^3 x^2 dx \, a d \tau.
\end{displaymath}
These will arrive during a time interval $dt_0 = a_0 d\tau _0= a_0 d
\tau$ and the fraction which will be detected by a detector of area
$A$ will be $A / 4 \pi a_0^2 x^2$. Thus on integrating back to a
distance equivalent to the time of equal energy and matter density,
which is as far back as is physically feasible and
summing over event types, the expected event rate will be:-
\begin{displaymath}
\Biggl \langle \frac{d{\cal N}}{dt_0} \Biggr \rangle =
\sum_{t}
A \int \limits_{0}^{c(\tau_0 - \tau_{eq})} N_t \, \langle n_t(x) \rangle \,
\frac{a^4}{a_0^3} \,dx.
\end{displaymath}
There is no point in integrating any further back because any cosmic
rays from such distances will not only have been heavily red-shifted
but also
severely affected by the medium through which they have travelled and
so would not be useful even in the unlikely event that they should
reach us intact.
Thus, on substituting for $\langle n_t (x) \rangle$ and $N_t$ and changing
variables to conformal time rather than $x$, one obtains:-
\begin{eqnarray}
\nonumber
\Biggl \langle \frac{d{\cal N}}{dt_0} \Biggr \rangle &=& \frac{Ac}{t_0^3}
\Biggl [ \frac{t_0}{t_{eq}} -1 \Biggr ]\Biggl \{ N_{ic}
 \frac{16 \chi \gamma^4}{81c^3} + N_{l} \frac{K \nu c}{(K-1)
\Gamma G \mu} \Biggr \}
\\ \nonumber
&+& \frac{4 Af \gamma ^2 }{27}
\biggl ( \frac{12 \epsilon ^2}{c^2w} \biggr ) ^{2/3}
\frac{1}{t_0^{7/3}}
\biggl \{ \biggl ( \frac{t_0}{t_{eq}}\biggr ) ^{1/3} -1 \biggr \}.
\end{eqnarray}

To gain an idea of the order of magnitude of the event rate, it is
necessary to substitute some of the values of the dimensionless
constants obtained from the simulations. Using very optimistic values
gives an upper limit on the eventual flux. At best, $\Gamma = 100$, $\gamma =
1$, $G \mu / c^4 = 10^{-6}$, $\chi = 0.1$, $w=10^{-31}$m, $\nu \approx
0.1$ and $K=10$.
The event rate is then:-
\begin{displaymath}
\Biggl \langle \frac{d{\cal N}}{dt_0} \Biggr \rangle = \frac{A}{c^2 t_0^3}
\Biggl [ \frac{t_0}{t_{eq}} -1 \Biggr ]
 \Biggl \{ 10^{-2} N_{ic} + 10^{5} N_{l} \Biggr \} + \frac{10^5 A}{(c^4 w^2
t_0^7)^{1/3}} \Biggl [ \biggl ( \frac{t_0}{t_{eq}} \biggr )^{1/3}
- -1 \Biggr ] .
\end{displaymath}
All of the numerical prefactors in this expression are dimensionless.
Notice that cusps are much more significant that loop collapses which
are in turn more significant than intercommutations. This
is to be expected as near us there is far more string in loops than in
long straight lengths and there is at least one cusp per loop in our
scheme. Thus, only the cusp term needs to be considered.

Adjusting the
numerical prefactors to obtain the event rate per year rather than per
second and neglecting the intercommutation and loop terms, the event
rate becomes:-
\begin{displaymath}
\Biggl \langle \frac{d{\cal N}}{dt_0} \Biggr \rangle = \frac{10^{12} A}{(c^4
w^2
t_0^7)^{1/3}} \Biggl [ \biggl ( \frac{t_0}{t_{eq}} \biggr )^{1/3}
- -1 \Biggr ] \approx 10^{-18} A \, \, {\rm year^{-1} }.
\end{displaymath}
This corresponds to an event rate of $10^{-10}$ per square kilometre
per century from cosmic strings. The flux of primary cosmic rays at
the highest energies is observed to be about one particle per square
kilometer per year \cite{sokolsky}.

An alternative comparison with
experiment may be obtained by the simple expedient of inserting a
redshift factor into the flux calculation and calculating an energy
flux, $F$. Omitting dimensionless factors of order unity, this turns
out to be:-
\begin{displaymath}
F = \Biggl \langle \frac{dE}{dt_0} \Biggr \rangle \approx A m_X \Biggl
( \frac{c^2 \epsilon^4}{w t_0^7} \Biggr ) ^{1/3}.
\end{displaymath}
If one then assumes that the energy spectrum is of the form
$\alpha E^{- \beta}$ above about $10^{17}$ eV then the differential
particle flux, which is the quantity normally plotted by
experimentalists,  is roughly $F/E^2$. The differential flux from
cosmic strings is several orders of magnitude lower than that observed.

In interpreting both of these comparisons, it should be borne in mind
that the predictions are an extreme upper limit on the flux due to
strings and do not include any form of attenuation. Further, they
include cosmic rays produced at the time of matter-energy equality
whereas a few hundred megaparsecs is probably far more reasonable. The
reason that the flux from cosmic strings is so small is that space is
so big and there are so few relevant events in it.

\section{Conclusions}

The flux of ultra-high-energy cosmic rays from gauge cosmic strings is
smaller than the observed fluxes by about ten orders of magnitude,
too small to be detectable with current technology. Thus any cosmic
rays observed above about
a joule are not due to cosmic strings. They can only be explained by a
modification of current ideas about acceleration mechanisms or some
entirely new physics.  The assumptions of previous authors which were
incorrect and led to their conclusion that cosmic strings could be
responsible for ultra-high-energy cosmic rays were:-
\begin{itemize}
\item The absolute amplitudes of the predicted spectra were found by
matching to experiment, thus assuming that strings are responsible for
ultra-high-energy cosmic rays, rather than from any knowledge of the
evolution of a cosmic string network. If this is done then there is a
tendency for the scenario to predict too large a proton flux at around
a few hundred  MeV, although it is possible to resolve this crisis
\cite{wolfendale} \cite{chi}.
\item The energy yielded by each event was sometimes expressed as a
fraction of a physical length, such as the length of a loop. This
makes scaling of the energy per event implicit in the calculation and
vastly overestimates the flux.
\end{itemize}
Also, this paper has improved on previous work by considering all of
the possible mechanisms for cosmic ray production at the same time and
in the same formalism.

It is possible that other types of defect could be a source for cosmic
rays. There are three which are cosmologically viable, viz.\ global
strings, textures and superconducting strings. All of these suffer to
some extent from the fact that they were invented more recently than
gauge strings and therefore their properties and evolution have been
less well studied than those of gauge strings. It is the authors'
conviction, however,
that these defects will also fail to provide an observable flux for
similar reasons to those for gauge strings. Also superconducting
strings will tend to lose much energy to relatively low frequency
synchrotron radiation which will not help.

\section{Acknowledgements}

The authors would like to thank the following for their conversations
and comments: A. Albrecht, D. Austin, P. Bhattacharjee, E. Copeland,
B. Drolias, M. Hindmarsh, J. Quenby, R.J. Rivers and E.P.S. Shellard.

\end{document}